\documentclass{article}
\usepackage{arxiv}
\usepackage{hyperref}
\usepackage{latexsym}
\usepackage{url}
\usepackage[utf8]{inputenc}
\usepackage{booktabs}
\usepackage{amsmath}
\usepackage{amsfonts}
\usepackage{amssymb}
\usepackage[inline]{enumitem}
\usepackage{enumitem}
\usepackage{mathbbol}
\usepackage[T1]{fontenc}
\usepackage{mathtools}
\usepackage{multirow}
\usepackage{natbib}
\usepackage{rotating}
\usepackage{subcaption}
\usepackage{graphicx}
\usepackage{etaremune}
\usepackage{authblk}
\usepackage{tikz}
\usepackage{tabularx}
\usepackage{arydshln}

\newcommand{\ie}{\emph{i.e.}}
\newcommand{\eg}{\emph{e.g.}}

\newcommand{\vs}{\emph{vs. }}

\title{Overview of the TREC 2023 deep learning track}

\author[1]{Nick Craswell}
\author[1]{Bhaskar Mitra}
\author[2,3]{Emine Yilmaz}
\author[2]{Hossein A.~Rahmani}
\author[4]{Daniel Campos}
\author[5]{Jimmy Lin}
\author[6]{Ellen M. Voorhees}
\author[6]{Ian Soboroff}
\affil[1]{Microsoft\\\texttt{\small \{nickcr, bmitra\}@microsoft.com}}
\affil[2]{University College London\\\texttt{\small \{emine.yilmaz, hossein.rahmani.22\}@ucl.ac.uk}}
\affil[3]{Amazon\\\texttt{\small eminey@amazon.co.uk}}
\affil[4]{Snowflake\\\texttt{\small daniel.campos@snowflake.com}}
\affil[5]{University of Waterloo\\\texttt{\small jimmylin@uwaterloo.ca}}
\affil[6]{NIST\\\texttt{\small \{ellen.voorhees, ian.soboroff\}@nist.gov}}

\begin{document}
\maketitle

\begin{abstract}
This is the fifth year of the TREC Deep Learning track.
As in previous years, we leverage the MS MARCO datasets that made hundreds of thousands of human-annotated training labels available for both passage and document ranking tasks.
We mostly repeated last year's design, to get another matching test set, based on the larger, cleaner, less-biased v2 passage and document set, with passage ranking as primary and document ranking as a secondary task (using labels inferred from passage).
As we did last year, we sample from MS MARCO queries that were completely held out, unused in corpus construction, unlike the test queries in the first three years.
This approach yields a more difficult test with more headroom for improvement.
Alongside the usual MS MARCO (human) queries from MS MARCO, this year we generated synthetic queries using a fine-tuned T5 model and using a GPT-4 prompt.

The new headline result this year is that runs using Large Language Model (LLM) prompting in some way outperformed runs that use the ``nnlm'' approach, which was the best approach in the previous four years.
Since this is the last year of the track, future iterations of prompt-based ranking can happen in other tracks.
Human relevance assessments were applied to all query types, not just human MS MARCO queries.
Evaluation using synthetic queries gave similar results to human queries, with system ordering agreement of $\tau=0.8487$.
However, human effort was needed to select a subset of the synthetic queries that were usable.
We did not see clear evidence of bias, where runs using GPT-4 were favored when evaluated using synthetic GPT-4 queries, or where runs using T5 were favored when evaluated on synthetic T5 queries. 

\end{abstract}
\section{Introduction}
\label{sec:intro}
At TREC 2023, we hosted the fifth TREC Deep Learning Track continuing our focus on benchmarking ad hoc retrieval methods in the large-data regime.
As in previous years~\citep{craswell2019overview, craswell2020overview, craswell2021overview, craswell2022overview}, we leverage the MS MARCO datasets~\citep{bajaj2016ms} that made hundreds of thousands of human annotated training labels available for both passage and document ranking tasks.
In addition, in 2021 we refreshed both the passage and the document collections which also led to a nearly $16$ times increase in the size of the passage collection and nearly four times increase in the document collection size.
In addition to evaluating ranking methods on the larger collections, the data refresh also aimed at providing additional metadata---\eg, passage-to-document mappings---that may be useful for ranking, as well as incorporating some fixes for known text encoding issues in previous versions of the datasets.
However, the significant increase in collection sizes in 2021 led to a corresponding increase in the number of relevant results in the collection per query and the existing judgment budget was exceeded before a reasonably complete set of these relevant results could be identified by the NIST judges.
This large number of relevant raised serious concerns about the test collection generated by the track in 2021, relating to reusability and also score saturation~\citep{voorhees2022too, craswell2021overview}.
Last year, these concerns were addressed via three specific changes:
\begin{enumerate*}[label=(\roman*)]
    \item Employing test queries that did not contribute to the MS MARCO corpus,
    \item employing NIST assessors to manually evaluate the relevance of retrieved results only for the passage ranking task and propagate the same labels to the source documents for the document ranking task, and finally,
    \item detection of near-duplicate passages and only judging one representative passage from each near-duplicate cluster with respect to the target query.
\end{enumerate*}
This year we continue to benchmark against these larger passage and document collections.

This year we evaluated the participant runs using both real user queries (our usual from MS MARCO) and synthetic queries from fine-tuned T5 and GPT-4.
Relevance assessments on different query types yielded consistent system orderings.
This year, we also received runs that employed large language models (LLMs) (run type ``prompt'') that seems to perform much better than fine-tuned LLMs on the ranking tasks.
Please see participant papers for more insights about what we learned this year.

\section{Task description}
\label{sec:task}

Similar to previous years, the Deep Learning Track in 2023 has two tasks: Passage ranking and document ranking. Participants were allowed to submit up to three official runs, and additional baseline runs, for each task. When submitting each run, participants indicated what external data, pretrained models and other resources were used, as well as information on what style of model was used.

The TREC Deep Learning Track has a focus on generating reusable test collections and analyzing reusability. In 2023, we primarily focused on rerunning the track under the same setup and using the same datasets as last year.
Hence, we again focused on the passage
ranking task as the primary task (while keeping the document ranking task as the secondary task) and mainly aimed
at constructing a sufficiently complete and reusable test collection for the passage ranking task. Labels inferred from
passage-level labels have then been used for the document ranking task.

Last year we proposed a new query sampling methodology with the intention of making the queries more difficult, to avoid the case where all runs have equally high performance and the evaluation is less discriminative.
This new method uses queries from the same sampling and annotation pipeline as standard MS MARCO queries. The pipeline samples Bing queries, uses a classifier to find queries that are answerable by a short passage, and since the classifier is imperfect the annotators can also reject a query as ``can't judge''. For consistency, we also eliminated queries where the judge did not select a passage, see Figure\ref{fig:ms_marco_hit}. The difference is that all our MS MARCO ranking datasets until last year were based on a 2018 version of the MS MARCO data with one million queries as described in a 2018 update of the MS MARCO paper \citep{bajaj2016ms}.\footnote{We note that the 2018 update of the paper \citep{bajaj2016ms} has an expanded author list, reflecting the expansion of the dataset to one million queries, which was planned by the original 2016 authors, and the addition of a ranking task, which was a new idea in 2018 not planned by the 2016 authors. The 2016 version and author list \citep{nguyen2016msmarco} reflect a preliminary release of the MS MARCO data, with 100 thousand queries and a natural language generation task.} The new annotations went through the same process, but after the one million query cutoff. This means they were not in the one million MS MARCO queries, their top-10 passages and URLs were not used to construct the MS MARCO passage and document corpora. It also means we do not have an evaluation using the MS MARCO sparse qrels and we did not filter out test queries where the sparse qrel failed to make it into the corpus. We expect queries from this new query sampling strategy to be more difficult and adopted the same approach this year.
In addition, we also explored evaluation using both real user queries and synthetic queries from fine-tuned T5 and GPT-4.

Sampling of human queries, generation of synthetic queries, and selection of which synthetic queries to include (see Section~\ref{sec:synthetic}) led to a set of 700 test queries used for both the passage ranking and document ranking tasks. 
The NIST assessment process allows assessors to reject a query before judging, and it may also be rejected after some judging. Queries can be rejected for being difficult to understand, or because there are too few or too many relevant results. The goal is to have a reusable test collection, meaning that a sufficiently complete set of relevant results have been identified that we can fairly evaluate a future system, that retrieves different relevant documents. This year there were $82$ queries that made it through the passage ranking assessment, and are used for evaluation. Document ranking runs on the same set of $82$ queries by propagating the passage labels to their source documents. 

Below we provide more detailed information about the document retrieval and passage retrieval tasks, as well as the datasets provided as part of these tasks.

\subsection{Passage ranking task}
The first task focuses on passage ranking, with two subtasks:
\begin{enumerate*}[label=(\roman*)]
    \item a full ranking and
    \item a top-$100$ reranking tasks.
\end{enumerate*}

In the full ranking subtask, given a query, the participants were expected to retrieve a ranked list of passages from the full collection based on the estimated likelihood of the passage containing an answer to the question.
Participants could submit up to $100$ passages per query for this end-to-end ranking task.

In the top-$100$ reranking subtask, $100$ passages per query were provided to participants, which were retrieved using Pyserini~\citep{lin2021pyserini}.
The reranking subtask allows all participants to start from the same starting point and to focus on learning an effective relevance estimator, without the need for implementing an end-to-end retrieval system.
It also makes the reranking runs more comparable, because they all rerank the same set of $100$ candidates.


Like last year, only passages were judged, with passage judgments subsequently propagated to documents to form the document relevance judgments.
In years before 2022, both documents and passages were judged independently for the track, so focusing assessing resources on only passages since 2022 effectively doubled the passage judgment budget.

The other major change was judging only a single element from a set of near-duplicate passages.
To effect this change, the passage corpus was clustered into classes of near-duplicate documents using the process at \url{https://github.com/isoboroff/dedupe}.
Each class had a single passage designated as the canonical passage for the class and the passage id of that passage was used as the class identifier.
The relevance label of the canonical passage with respect to a query was propagated to all the other passages in the same class.

The track received 35 submissions to the passage ranking task, 14 of which were baseline runs.
All 35 submitted runs contributed to the initial judgment pools. The judging process was similar to last year but without the collection subsets.

Judgments were collected on a four-point scale:
\begin{etaremune}[start=3]
    \item \textbf{Perfectly relevant:} The passage is dedicated to the query and contains the exact answer.
    \item \textbf{Highly relevant:} The passage has some answer for the query, but the answer may be a bit unclear, or hidden amongst extraneous information.
    \item \textbf{Related:} The passage seems related to the query but does not answer it.
    \item \textbf{Irrelevant:} The passage has nothing to do with the query.
\end{etaremune}
For metrics that binarize the judgment scale, we map passage judgment levels 3,2 to relevant and map passage judgment levels 1,0 to irrelevant.

\subsection{Document ranking task}

Similar to the passage ranking task, the document ranking task focuses on two subtasks:
\begin{enumerate*}[label=(\roman*)]
    \item Full ranking and
    \item top-$100$ reranking.
\end{enumerate*}

The full ranking subtask models the end-to-end retrieval scenario, documents can be retrieved from the full document collection provided and the runs are expected to rank documents based on their relevance to the query. 

Similar to passage ranking, in the document reranking subtask, participants were provided with an initial ranking of $100$ documents, giving all participants the same starting point. The $100$ documents provided to the participants were generated using Pyserini~\cite{lin2021pyserini}. Participants were expected to rerank the 100 documents based on their estimated likelihood of containing an answer to the query.

Instead of collecting additional judgments for the document ranking task, we used passage judgments to infer judgments for documents: For each document we first identified the passages that were judged from within that document when collecting judgments for the passage ranking task, where all duplicates of a judged passage are assumed to have the same relevance judgment as the judged passage. If a document contains multiple passages with associated relevant judgments, we use the max judgment across all the passages to infer the final relevance judgment for the document. Previous work has shown that such an approach results in reasonable quality relevance judgments~\citep{Zhijing2019passage2doc}, and our study on the 2021 test collections further validated this~\citep{craswell2021overview}. 

Different from the passage ranking task, for document ranking metrics that use binary judgments we map document judgment levels 3,2,1 to relevant and map document judgment level 0 to irrelevant.

\section{Datasets}
\label{sec:data}

This year we again used the MS MARCO v2 dataset. To understand how the new dataset differs from the old, we will first describe the natural language generation dataset from 2016 and the v1 ranking data from 2018.

\begin{figure}
    \centering
    \includegraphics[width=0.75\linewidth]{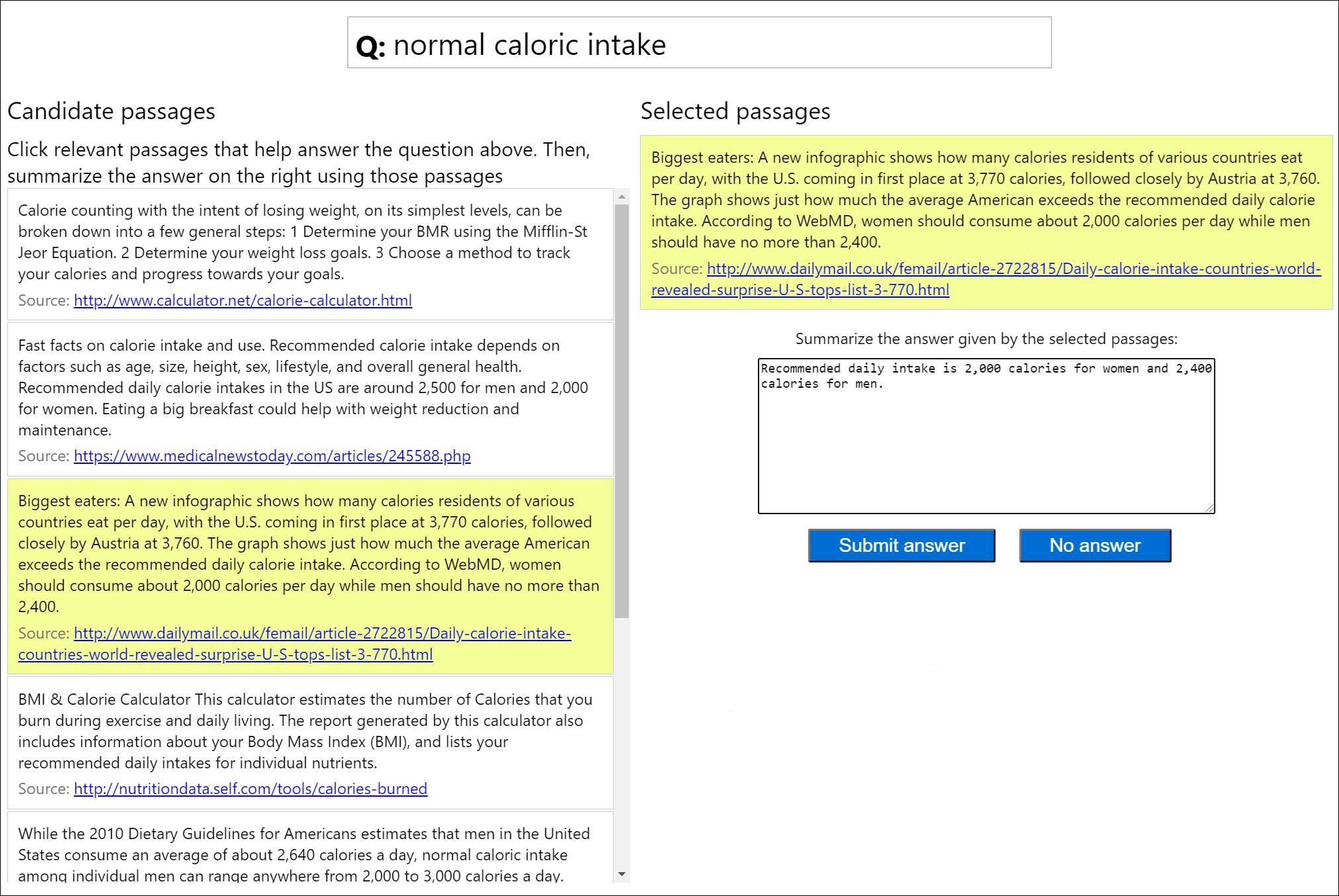}
    \caption{Crowd task used to generate the original MS MARCO natural language generation leaderboard. This same crowd data was later adapted to become the MS MARCO ranking tasks.}
    \label{fig:ms_marco_hit}
\end{figure}

\paragraph{MS MARCO natural language generation dataset.}
The original MS MARCO dataset was for a natural language generation task, rather than a ranking task. It processed one million queries, using a crowd task as shown in Figure~\ref{fig:ms_marco_hit}. The crowd worker would read the query, consider up to ten passages related to the query, decide if the passages could be used to answer the query and if answerable write an answer to the query in their own words. For each answerable question the crowd workers provided a non-extractive answer and an annotation of which passages they used to generate their answer. There was substantial quality work with the crowd workers to ensure quality and the crowd workers spent an average of 2.5 minutes on each annotation. The million queries were drawn from actual user queries to Bing. The ten results were generated by a Bing passage retrieval and ranking system. The queries were filtered before being annotated to remove any adult or offensive queries and any non-English queries. Moreover, further filtering was performed to ensure that the queries came from the $10-20\%$ of English queries that were detected as potentially being answerable with a short passage. Although the filter may be imperfect, the intention was to exclude navigational queries (such as [youtube]), queries that require a longer answer (such as [beef wellington recipe]) and queries that aim to complete some transaction (such as [buy xbox live]). We note that about 35\% of the queries could not be answered using the ten passages, in which case the crowd worker would indicate No answer, and one part of the original MS MARCO challenge was to predict which queries were answerable.

\paragraph{MS MARCO ranking v1 datasets.}
The MS MARCO passage and document ranking v1 datasets are used in the current MS MARCO leaderboards~\citep{lin2022fostering, craswell2021ms, lin2021significant} and in TREC 2019 and TREC 2020.

To generate the v1 passage ranking data, we took the union of the top ten passage lists for the one million queries, giving us 8.8 million distinct passages. For queries that were answerable, we used the crowd judge annotation for selected passages as a positive qrel. This gives us highly incomplete qrels, as noted in the original description \citep{bajaj2016ms}. We should in no way expect the positive qrel to be the ``best answer''. We found that training and evaluating using these sparse qrels gives us results that are quite correlated with results using much more comprehensive NIST judgments \citep{craswell2019overview, craswell2020overview}. Further study is needed to understand why this works, but we suspect it's important that the qrel is selected from a Bing ranking that has access to information that's unavailable to TREC participants, such as billions of past queries. This means the selected qrel is not biased towards some existing academic approach such as BM25. For each query that has a qrel, we generated a BM25 top-1000 for use in a reranking task and also allowed fullrank from the 8.8 million passages. We used the same split as in the QnA task: training ($80\%$), dev ($10\%$) and eval ($10\%$).

To generate the v1 document ranking data, we collected the corresponding urls for which the passages were extracted. Using these 3.5 million URLS, we obtained the associated document title and body corresponding to the ranking qrels. It is worth noting that the original passages were extracted between January 2016 and February 2018 while the full documents were extracted in March of 2018 and as a result only 3.2 million URLs were still able to be successfully extracted. From these documents, a \textit{clean} form was extracted where the body text had the HTML removed and focused on the main content of the page, removing web-page boilerplate such as navigation menus. Since we extracted the document text more than a year later than the passage data and used a completely different document parsing and processing pipeline (which unfortunately had character set processing issues) there was a chance that some pages that had a relevant passage no longer existed, no longer contained the passage, or even had the section of text with the passage accidentally removed as boilerplate. These are all realistic things to happen in a real-world application, where the document corpus is constantly changing, we do not wish to throw away our old relevance labels, and indeed we may not have budget to generate new labels. Doing a better job of generating a clean dataset using old labels is what we have now done in generating the v2 data. Qrels for the document task were assigned by assuming that a relevant passage qrel transfers to the document level as a positive document qrel. We generated top-100 document rankings using Indri, for use in a reranking task and also allowed fullrank from the 3.2 million documents.

The v1 data had several problems. The corpus was generated based on the queries, such that each passage and each document is in the corpus due to one of our million original queries. For each document in the corpus there may only be one passage in the passage dataset (and on average 2.8 passages per document), but that passage was identified by Bing in relation to one of the MS MARCO queries, possibly a test query. This is unrealistic, since a real system would be able to generate many candidate passages per document, and would not know what the test queries will be ahead of time. Therefore, we had to forbid participants from considering the passage-document mapping. The document dataset had several problems with character sets and missing whitespace.

\paragraph{MS MARCO ranking v2 datasets.}
The MS MARCO passage and document ranking v2 datasets were used for the first time in TREC 2021. The goal of the v2 dataset was to increase the scale and introduce a wider variety of documents such that not all documents were relevant to at least some query.

While the v1 data started with passages and was expanded to documents, the v2 data is document native. It begun by identifying documents based on the source urls of the v1 dataset. Of the original 3.5 million MS MARCO URLs, we were able to still find content for 2.7 million. We added an additional 9.2 million documents, selected to be the kind of documents that had useful passages of text in past Bing queries, giving a total of 11.9 million documents. For each document we ran a query-independent proprietary algorithm for identifying promising passages, and selected the best non-overlapping passages, giving on average 11.6 passages per document. This gives us our 138 million passages in the v2 passage corpus. We mapped the document qrels at the URL level, for training, dev and eval. The chance that the document is no longer relevant to the query, which also was a concern in v1 data, is now increased since the document content was extracted at a later date. We can consider how big this problem is by analyzing the disagreement rate between MS MARCO qrels and NIST qrels (in v1 and v2), and seeing whether training on MS MARCO qrels yields improved NIST NDCG on the test set. For mapping passage qrels, we required that the passage comes from the same URL as the original passage, and has sufficient text similarity to the positive passage text from v1. 

It is now possible for participants to use the passage-document mapping in participation, for example by considering document information in passage ranking, passage information in document ranking, and so on. Using a larger corpus prevents participants from proposing completely unscalable ranking approaches. The new dataset has fewer character encoding and whitespace issues, and could form the basis for future tasks that include some elements of additional document processing, such as extracting even shorter (phrase) answers.

\section{Results and analysis}
\label{sec:result}

\paragraph{Submitted runs}
A total of six groups participated in the TREC 2023 Deep Learning Track. Among them, five groups participated in the passage ranking task, and two groups participated in the document ranking task.
Similar to previous years, we submission of baseline runs. 
This adds diversity to the relevance assessment pools and encourages a participating group to compare a new approach with a well-understood existing approach.
Across all groups, we received a total of $40$ run submissions, including $35$ passage ranking runs and five document ranking runs.
Table~\ref{tbl:runs-by-type} summarizes the submissions statistics for this year's track.

\begin{table}
    \centering
    \caption{TREC 2023 Deep Learning Track run submission statistics.}
    \begin{tabular}{lrr}
    \hline
    \hline
        & \textbf{Passage ranking} & \textbf{Document ranking} \\
        \hline
        Number of groups & 5 & 2 \\
        Number of total runs & 35 & 5 \\
        Number of baseline runs & 14 & 1 \\
        Number of runs w/ category: prompt & 19 & 1 \\
        Number of runs w/ category: nnlm & 14 & 4 \\
        Number of runs w/ category: nn & 0 & 0 \\
        Number of runs w/ category: trad & 2 & 0 \\
        Number of runs w/ category: rerank & 4 & 0 \\
        Number of runs w/ category: fullrank & 31 & 5 \\
        \hline
        \hline
    \end{tabular}
    \label{tbl:runs-by-type}
\end{table}

This year we asked participants to self-classify each of their runs under the following four categories:
\begin{itemize}
    \item trad: No neural representation learning---\eg, classical learning to rank, PRF, and BM25
    \item nn: Representation learning with text as input, but not using a pre-trained model
    \item nnlm: Using a pre-trained model in any part of the pipeline---\eg, neural document expansion and BERT-style reranking
    \item prompt: Using a large language model with prompt in any part of the pipeline
\end{itemize}

Last year we were interested to see if there were runs that were doing a single stage of retrieval with dense retrieval. Such a system converts the query into a vector, retrieves document vectors based on vector comparison, and then presents the ranking based on vector comparison to the user. Such a system could have the advantage of simplicity and low query latency, compared to a system that has a candidate generation stage (using dense retrieval and/or traditional indexing) followed by one or mpre phases of reranking. However, we did not find that participants were highly successful with a ``dense retrieval only'' approach, and also that some top runs did not use dense retrieval at all.

This year we were interested to see if people would use a prompt-based LLM approach in their runs. There is some chance that LLM approaches can outperform other forms of ranking, giving us another step above the ``nnlm'' runs that we studied in the first four years of the track. We expected prompt-based approaches could perform well, and the prompt-based component in the retrieval system may not need to fine-tune on MS MARCO training data, but we also expected MS MARCO fine tuning would be required at some stage in the ranking stack.

To understand this we can consider the performance of systems with:
\begin{itemize}
    \item Run type ``prompt'', and
    \item Training data: none, other, marco. 
\end{itemize}
We anticipated good performance from ``prompt'', but in a ranking stack with LLM prompting we expected that MS MARCO training data (``marco'') might still be needed, for example to fine-tune early dense and sparse retrieval stages.

\paragraph{Overall results}
Table~\ref{tab:passage_ranking} and Table~\ref{tab:document_ranking} present a standard set of relevance quality metrics for document and passage ranking runs, respectively, as we have reported for the track in previous years.
The reported metrics include Normalized Discounted Cumulative Gain (NDCG)~\citep{JK2002}, and Average Precision (AP)~\citep{voorhees2005trec}. These are all computed using NIST judgments, since this year's test queries do not have the sparse judgments that we used in the first three years.

In subsequent discussions, we employ NDCG@10 as our primary evaluation metric to analyze ranking quality produced by different methods.
To analyze how different approaches compare beyond just the relevance of top-ranked results, we use NDCG@100 and Average Precision.

\begin{table}[]
\caption{Summary of results for passage ranking runs. Baseline submissions marked with (b).}
\scriptsize
\centering
\begin{tabular}{lllllrrr}
\toprule
                          run &        group &  subtask & run type & training data &  ndcg@10 &  ndcg@100 &     ap \\
\midrule
           naverloo-rgpt4 (b) &       h2oloo & fullrank &   prompt &         marco &   0.6994 &    0.5370 & 0.3382 \\
              naverloo-frgpt4 &       h2oloo & fullrank &   prompt &         marco &   0.6899 &    0.5362 & 0.3300 \\
       naverloo\_fs\_RR\_duo (b) &       h2oloo & fullrank &   prompt &         marco &   0.6585 &    0.5291 & 0.3130 \\
                    cip\_run\_2 &          CIP & fullrank &   prompt &         marco &   0.6558 &    0.4927 & 0.2862 \\
                    cip\_run\_1 &          CIP & fullrank &   prompt &         marco &   0.6558 &    0.4927 & 0.2862 \\
                    cip\_run\_3 &          CIP & fullrank &   prompt &         marco &   0.6185 &    0.4823 & 0.2714 \\
                    cip\_run\_4 &          CIP & fullrank &   prompt &         marco &   0.6137 &    0.4839 & 0.2721 \\
                    cip\_run\_6 &          CIP & fullrank &   prompt &         marco &   0.6078 &    0.4810 & 0.2677 \\
                    cip\_run\_7 &          CIP & fullrank &   prompt &         marco &   0.6075 &    0.4813 & 0.2676 \\
                    cip\_run\_5 &          CIP & fullrank &   prompt &         marco &   0.6074 &    0.4776 & 0.2656 \\
           naverloo\_fs\_RR (b) &       h2oloo & fullrank &     nnlm &         marco &   0.5972 &    0.5143 & 0.2844 \\
     naverloo\_bm25\_splades\_RR &       h2oloo & fullrank &     nnlm &         marco &   0.5891 &    0.5066 & 0.2811 \\
             uogtr\_b\_grf\_e\_gb &        uogTr & fullrank &   prompt &         marco &   0.5489 &    0.4386 & 0.2314 \\
               uogtr\_qr\_be\_gb &        uogTr & fullrank &   prompt &         marco &   0.5488 &    0.4350 & 0.2315 \\
                  uogtr\_be\_gb &        uogTr & fullrank &     nnlm &         marco &   0.5451 &    0.4326 & 0.2285 \\
                  uogtr\_se\_gb &        uogTr & fullrank &     nnlm &         marco &   0.5394 &    0.4443 & 0.2401 \\
             naverloo\_bm25\_RR &       h2oloo & fullrank &     nnlm &         marco &   0.5378 &    0.4115 & 0.2070 \\
                uogtr\_b\_grf\_e &        uogTr & fullrank &   prompt &         marco &   0.5376 &    0.3987 & 0.1996 \\
                 uogtr\_se (b) &        uogTr & fullrank &     nnlm &         marco &   0.5364 &    0.4340 & 0.2348 \\
                  uogtr\_qr\_be &        uogTr & fullrank &   prompt &         marco &   0.5316 &    0.3938 & 0.1994 \\
                 uogtr\_be (b) &        uogTr & fullrank &     nnlm &         marco &   0.5227 &    0.3873 & 0.1940 \\
              naverloo\_fs (b) &       h2oloo & fullrank &     nnlm &         marco &   0.5045 &    0.4274 & 0.2116 \\
    splade\_pp\_self\_distil (b) &       h2oloo & fullrank &     nnlm &         marco &   0.4768 &    0.3942 & 0.1960 \\
         slim-pp-0shot-uw (b) &       h2oloo & fullrank &     nnlm &         marco &   0.4762 &    0.3658 & 0.1773 \\
splade\_pp\_ensemble\_distil (b) &       h2oloo & fullrank &     nnlm &         marco &   0.4730 &    0.3905 & 0.1924 \\
                  uogtr\_s (b) &        uogTr & fullrank &     nnlm &         marco &   0.4706 &    0.3910 & 0.1925 \\
             bm25\_splades (b) &       h2oloo & fullrank &     nnlm &         marco &   0.4590 &    0.4048 & 0.1886 \\
          agg-cocondenser (b) &       h2oloo & fullrank &     nnlm &         marco &   0.4562 &    0.3563 & 0.1776 \\
           uot-yahoo\_rankgpt4 &    uot-yahoo &   rerank &   prompt &         other &   0.3927 &    0.2651 & 0.1171 \\
              WatS-LLM-Rerank & UWaterlooMDS &   rerank &   prompt &         marco &   0.3706 &    0.2608 & 0.1069 \\
          uot-yahoo\_rankgpt35 &    uot-yahoo &   rerank &   prompt &         other &   0.3376 &    0.2499 & 0.0969 \\
                uogtr\_dph (b) &        uogTr & fullrank &     trad &          none &   0.2825 &    0.2324 & 0.0840 \\
          WatS-Augmented-BM25 & UWaterlooMDS & fullrank &   prompt &         marco &   0.2696 &    0.2108 & 0.0755 \\
       uot-yahoo\_LLMs-blender &    uot-yahoo &   rerank &   prompt &         other &   0.2630 &    0.2258 & 0.0758 \\
            uogtr\_dph\_bo1 (b) &        uogTr & fullrank &     trad &          none &   0.2377 &    0.1589 & 0.0600 \\
\bottomrule
\end{tabular}

\label{tab:passage_ranking}
\end{table}

\begin{table}[]
\caption{Summary of results for document ranking runs. Baseline submission marked with (b).}
\scriptsize
\centering
\begin{tabular}{lllllrrr}
\toprule
                    run &     group &  subtask & run type & training data &  ndcg@10 &  ndcg@100 &     ap \\
\midrule
      D\_naverloo-frgpt4 &    h2oloo & fullrank &   prompt &         marco &   0.6893 &    0.6349 & 0.4189 \\
D\_naverloo\_bm\_splade\_RR &    h2oloo & fullrank &     nnlm &         marco &   0.5961 &    0.5994 & 0.3793 \\
     D\_naverloo\_bm25\_RR &    h2oloo & fullrank &     nnlm &         marco &   0.5587 &    0.5142 & 0.3131 \\
     D\_bm25\_splades (b) &    h2oloo & fullrank &     nnlm &         marco &   0.4936 &    0.5145 & 0.2976 \\
              colbertv2 & InfoSense & fullrank &     nnlm &         marco &   0.3133 &    0.2839 & 0.1297 \\
\bottomrule
\end{tabular}\label{tab:document_ranking}
\end{table}

\begin{figure}
  \center
  \begin{subfigure}{.49\textwidth}
    \includegraphics[width=\textwidth]{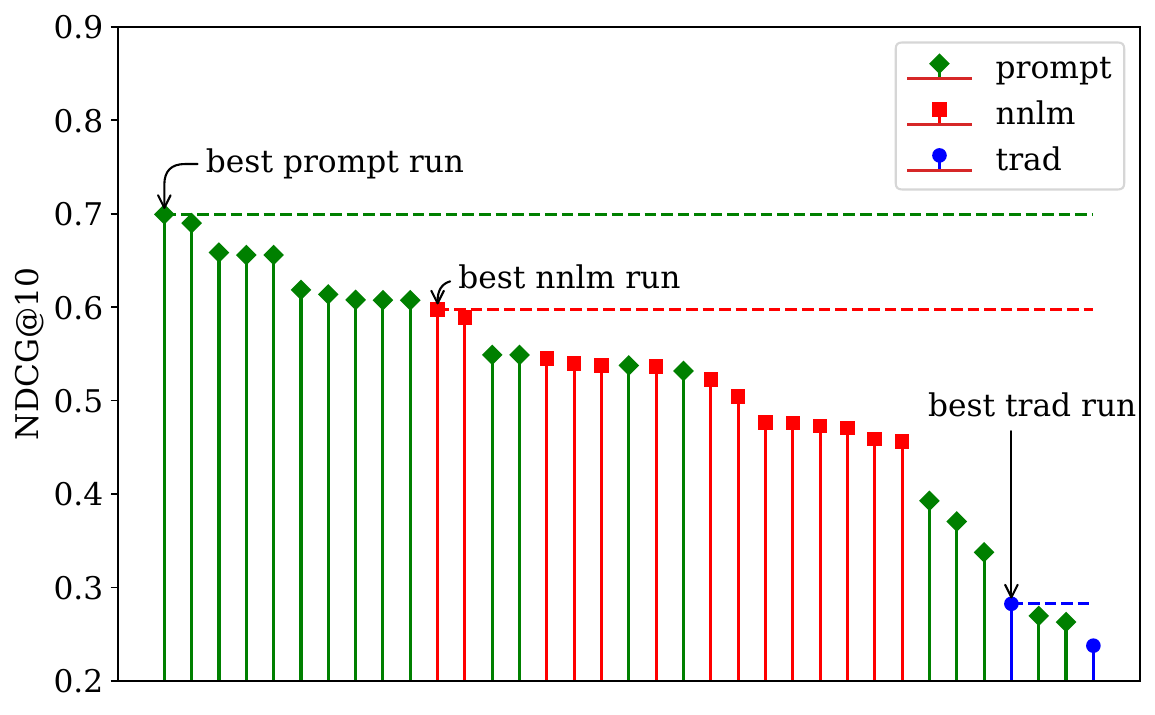}
    \caption{Passage ranking task}
    \label{fig:model-task-passages-stem-by-model-type}
  \end{subfigure}
  \hfill
  \begin{subfigure}{.49\textwidth}
    \includegraphics[width=\textwidth]{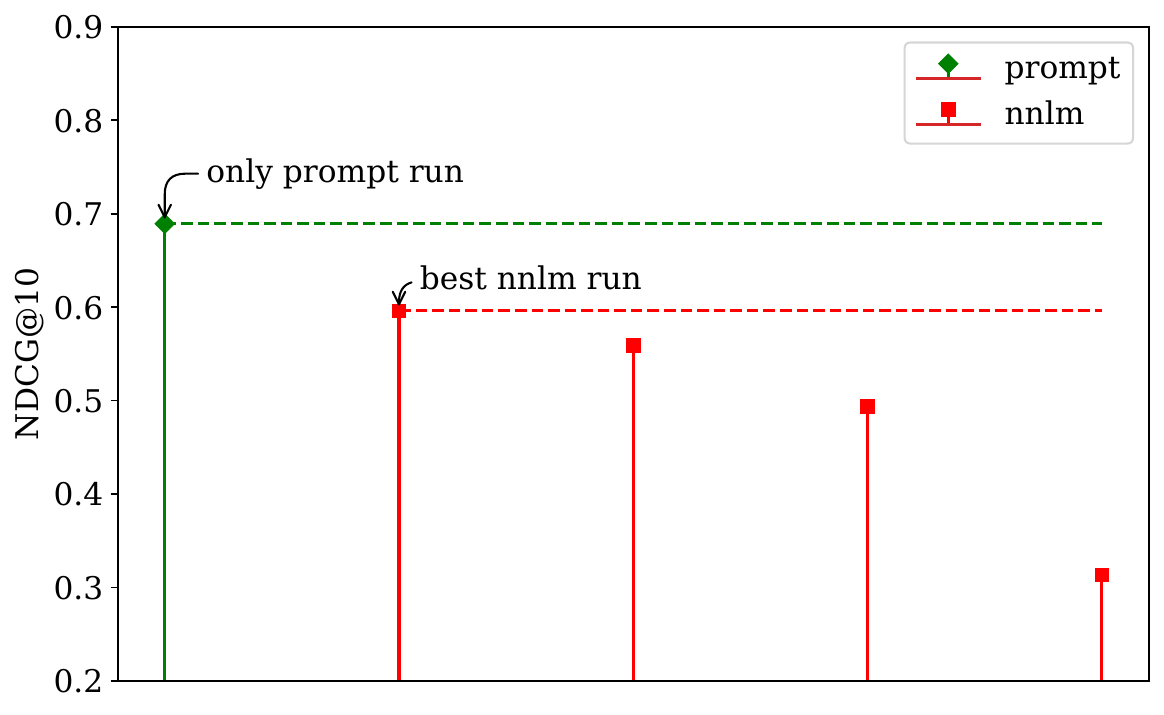}
    \caption{Document ranking task}
    \label{fig:model-task-docs-stem-by-model-type}
  \end{subfigure}
  \caption{NDCG@10 results by run type. As in the previous two years, ``nnlm'' runs continue to outperform over ``trad'' runs for both tasks.}
  \label{fig:model-stem-by-model-type}
\end{figure}

\paragraph{Prompt methods}
Figure~\ref{fig:model-stem-by-model-type} summarizes the evaluation results by run type---\ie, comparing ``prompt'' \vs ``nnlm'' \vs ``trad'' runs.
In previous years, ``nnlm'' methods dramatically outperformed ``trad'' runs, meaning that learning a representation from training data was very helpful. This year we see another gap starting to form, with ``prompt'' runs outperforming runs that did not use an LLM with prompting. We note that there are runs using very expensive language models such as GPT-4, and this was certainly not used to process the entire corpus of 138 million passages for each query. We can refer to the participant papers to see how the expensive model was used, but we might expect that trad and nnlm methods were used to retrieve a small number of candidates from the 138 million passage corpus, then the LLM approach was applied to a few results per query to improve their ranking. So, combining methods studied in previous years with a prompt-based LLM approach, and therefore classifying the run as ``prompt'', leading to better NDCG@10 than non-prompt approaches.

\begin{figure}
\includegraphics[width=\textwidth]{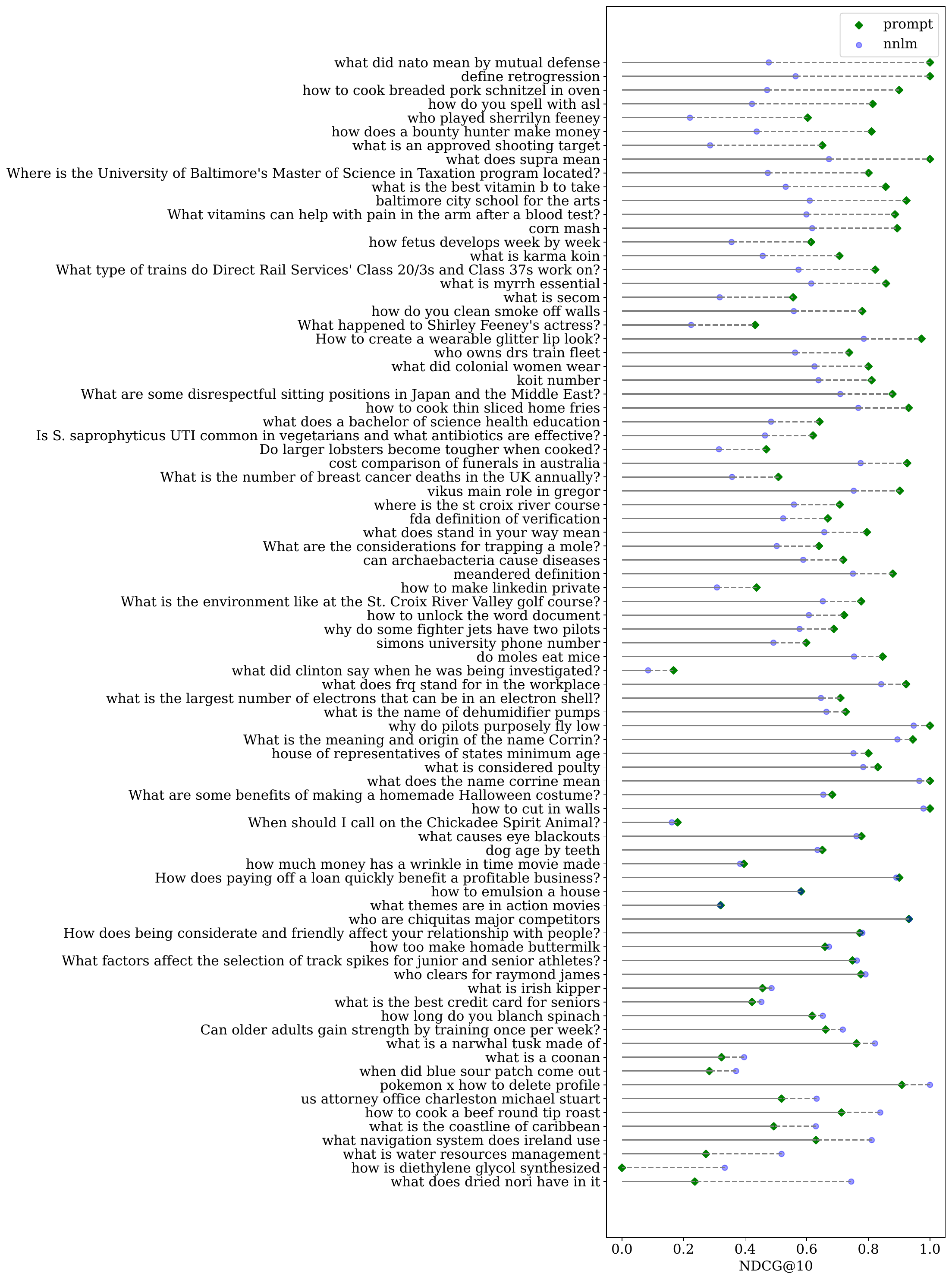}
\caption{Comparison of the best ``prompt'' and ``nnlm'' runs on individual test queries for the passage ranking task. Queries are sorted by difference in mean performance between ``prompt'' and ``nnlm'' runs. Queries on which ``prompt'' wins with large margin are at the top.}
\label{fig:model-task-passages-bar-per-query}
\end{figure}
\begin{figure}
\includegraphics[width=\textwidth]{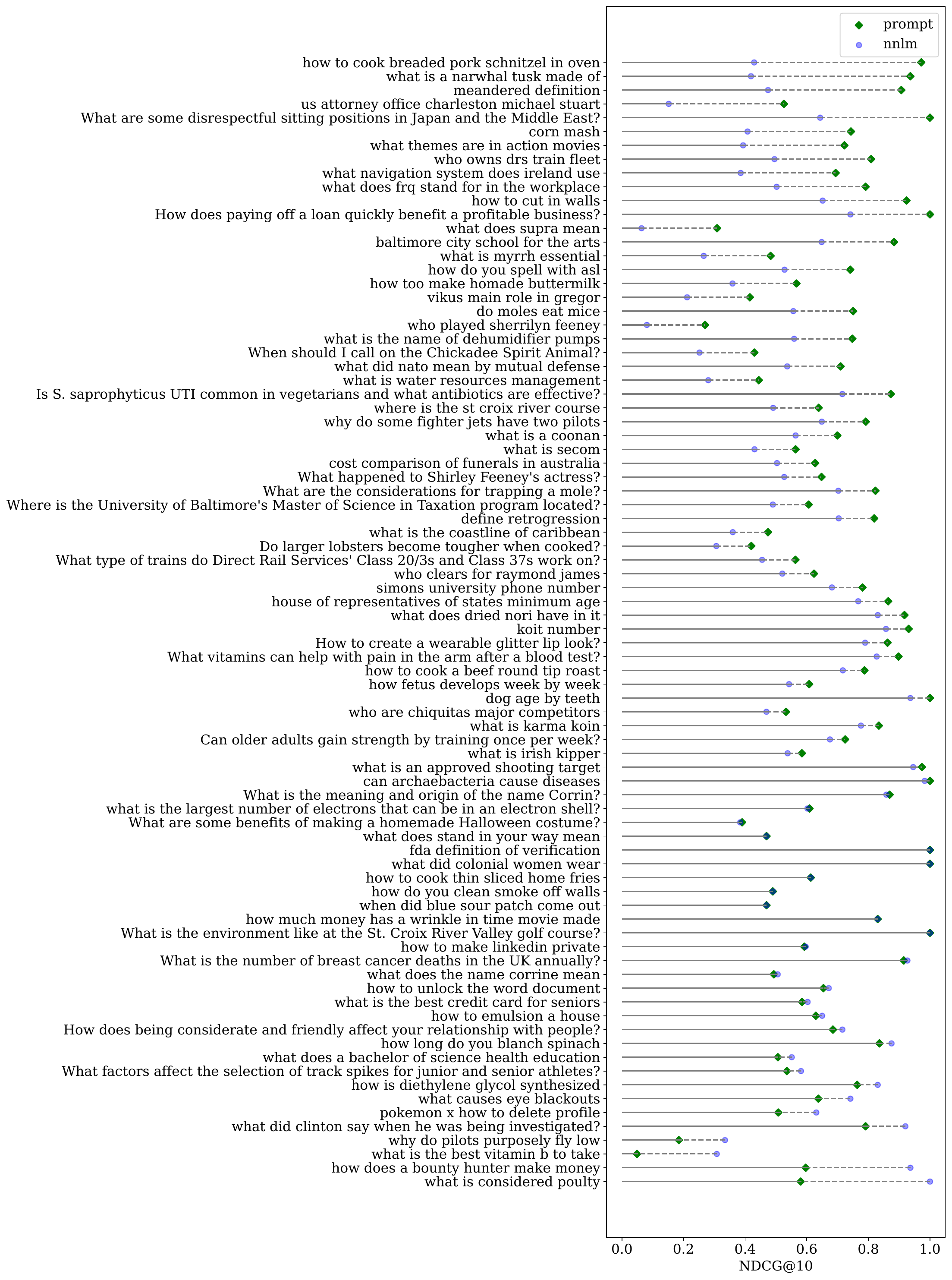}
\caption{Comparison of the best ``prompt'' and ``nnlm'' runs on individual test queries for the document ranking task. Queries are sorted by difference in mean performance between ``prompt'' and ``llm'' runs. Queries on which ``prompt'' wins with large margin are at the top.}
\label{fig:model-task-docs-bar-per-query}
\end{figure}

Figure~\ref{fig:model-task-passages-bar-per-query} and \ref{fig:model-task-docs-bar-per-query} show a query-level comparison between the best ``prompt'' and ``nnlm'' runs for the passage and the document ranking tasks, respectively.
The best ``prompt'' run outperforms the best ``nnlm'' on on the majority of queries, reminiscent of previous years where ``nnlm'' outperformed ``trad'' on the majority of queries. 
The same holds true for the document ranking task.

\paragraph{MS MARCO training data}
We see good performance for ``prompt'' approaches. Typically such LLM-based prompting approaches use a few-shot or zero-shot approach, meaning that they need few or none of the hundreds of thousands of MS MARCO training labels. It is not impossible to fine-tune an LLM using MS MARCO, and some might consider the few-shot approach to be using the MS MARCO labels. However, our interest is how well a prompt approach can work (with no fine tuning, but few-shot is reasonable to include), in a system that has no use of MS MARCO for fine tuning any dense retrieval or ranking stages.

Table~\ref{tab:passage_ranking} shows that most top runs using the ``prompt'' approach did self report as using the MS MARCO training data. We can refer to participant papers in more detail, but the overall results seem to indicate that the ``prompt'' approach does not obviate the need for fine tuning some other parts of the model.

\paragraph{Effect of near duplicates}
This year, as described in last year's overview \citep{craswell2022overview} we use the expanded qrels. This means that we avoid expending judging effort on judging many copies of the same near-dupe passages, and the official qrels instead map the relevance label of one passage in the cluster to the other passages. Evaluation should be done using trec\_eval on the official expanded qrels, which should be the only qrels available.

\section{Synthetic Query Analysis}
\label{sec:synthetic}
One of the main goals of the track has been to build reliable and reusable test collections, which highly depend on the queries included in the collection. Since obtaining real user queries may not always be possible, we also analysed the feasibility of using synthetic queries in test collection construction. Hence, in this year's track, in addition to real queries, we also used a set of synthetic queries generated using language models. 

To generate the synthetic queries, we first sampled 1000 passages at random from the v2 passage corpus\footnote{\url{https://msmarco.blob.core.windows.net/msmarcoranking/msmarco_v2_passage.tar}}. 
In the MS MARCO task, a retrieved passage should make sense on its own, when appearing on the search results page without the rest of the document's context.
Examples of a bad passage could be one that talks about a person without saying the person's name (``she then completed her first novel\ldots'''), or using terms that might not be clear once the passage is removed from its document context (``we eventually chose the second solution, since it gave a good balance of accuracy and efficiency'').
To identify passages that might not be good stand-alone search results we ran each of the 1000 passages through a GPT-4 prompt as shown in Table~\ref{tbl:passage_score_prompt}.
We ran each of the 1000 passages as a GPT-4 prompt with temperature=0.8, to get a query-independent passage selection score.
Since this is a rough one-off filter, we didn't repeat in the case of some error or malformed output, which eliminated 8.9\% of passages.
The following example is one of passages that with its rating and reasoning generated by GPT-4, with a quality rating of 30:
\begin{small}
\begin{verbatim}
  pid=msmarco_passage_12_127299414
  passage=Set-Acl C:\\Users\\myusername -AclObject $inheritance. This will take likely a few
    minutes if you have a lot of files like did. However, it seemed to reset the permissions in
    that folder to inherit from C:\\Users\\myusername in all child folders.
  rating=30
  reasoning=The passage does not clearly state the main topic or question it addresses, and it
    lacks necessary background information for the reader to understand the context. The passage
    also uses technical terms that are not explained or familiar to a general audience.
\end{verbatim}
\end{small}
Here is an example passage with a score of 90:
\begin{small}
\begin{verbatim}
  pid=msmarco_passage_65_620984460
  passage=GRE Test Scores. Someone who signs up for the GRE general test can expect to receive 
    three scores after successfully completing the test, including scores in verbal reasoning, 
    quantitative reasoning and analytical writing. Verbal and quantitative scores range from 130 
    to 170, and these scores are always a whole number.
  rating=90
  reasoning=The passage clearly states the main topic, which is GRE Test Scores. It explains the
    three scores a test-taker can expect to receive: verbal reasoning, quantitative reasoning, 
    and analytical writing. It also provides the score range for verbal and quantitative 
    sections. However, the passage does not mention the score range for analytical writing, 
    which would make it even more informative.
\end{verbatim}
\end{small}
Here is another example passage with a score of 90, although for our our purposes it still might not be useful, so the GPT-4 filter is not perfect:
\begin{small}
\begin{verbatim}
  pid=msmarco_passage_21_28155342
  passage = Andrew Stein (born 1945), American politician. Andy Stein, American saxophone and 
    violin player. Arthur Stein (born 1950), American professor of political science. Aurel
    Stein (1862–1943), Hungarian-British archaeologist.
  rating=90
  reasoning=The passage clearly states the main topic it addresses by providing a list of 
    individuals with the surname 'Stein' and their professions. The passage provides background 
    information on each individual, including their birth year and occupation. However, the 
    passage does not provide detailed information about their work or accomplishments, which 
    might be relevant to the intended audience.
\end{verbatim}
\end{small}
We filtered low quality passages with passage quality scores less than 50, removing $14.6\%$ passages. The remaining passages were seeds for this year's synthetic queries.

\begin{table}[!h]
    \caption{Prompts used in our synthetic query generation on randomly selected passages to generate a quality score for each passage. At the placeholders \{$passage$\} the actually sampled passage is included.}
    \label{tbl:passage_score_prompt}
        \vspace{1em}
        \begin{tabularx}{\linewidth}{X}
            \toprule
            \bf Passage Quality Score Generation\\
            \midrule
            Below is an instruction that describes a task, paired with an input that provides further context. Write a response that appropriately completes the request. \\\\
            
            \#\#\# Instruction: \\\\
            
            You are part of a search engine which helps users find relevant context passages given their queries. As part of the index generation process we are evaluating the quality of passages that could occur on a search results page.
            
            In order to meet user needs it is important to ensure the context passages state clearly what they are talking about, for example, if a passage is reviewing a product, it must say what the product is. Otherwise there is a risk that we are taking the review of one product, but the person who is reading it thinks it is talking about a different product, because the search engine accidentally picked the wrong passage. \\\\
            
            Read the context passage and rate it on a scale of 0 to 100, where 0 means that the passage is very unclear and confusing due to lack of context, and 100 means that the passage is very clear and informative due to providing sufficient context. Please include your reasoning and respond in well formed json. \\\\
            
            Consider the following factors when rating the passage:\\
            1. Does the passage state the main topic or question that it addresses?\\
            2. Does the passage provide any background information or definitions that are necessary to understand the main points?\\
            3. Does the passage avoid using jargon or technical terms that are not explained or familiar to the intended audience? \\\\
            
            \#\#\# Input: \\\\

            \{$passage$\} \\\\

            \#\#\# Response: \\
            
            

            \bottomrule
        \end{tabularx}

\end{table}

\begin{table}[!h]
    \caption{Prompts used in our synthetic query generation on randomly selected passages to generate a query for each passage. At the placeholders \{$passage$\} the actually sampled passage is included.}
    \vspace{1em}
    \begin{tabularx}{\linewidth}{X}
        \toprule
        \bf Query Generation Based on Passage\\
        \midrule
        Below is an instruction that describes a task, paired with an input that provides further context. Write a response that appropriately completes the request. \\\\
        
        \#\#\# Instruction: \\\\
        
        You are part of a search engine that helps users find relevant context passages given their queries.
        To improve the relevance of retrieval methods, artificial queries must be generated.
        Read the context passage and write a query that a user would issue if they would seek to access the passage mentioned below. Provide reasoning for your generation. \\\\
        
        Consider the following factors when generating the query:\\
        1. Does this passage fully answer the query?\\
        2. Does this query seek information that could only be answered by this given passage?\\\\

        Respond in JSON where the generated query has the key "query" and the reasoning has the key "reasoning."\\\\
        
        \#\#\# Input: \\\\

        context passage:\{$passage$\} \\\\

        \#\#\# Response: \\
        \bottomrule
    \end{tabularx}
\end{table}

We then generated queries using two methods: The first method uses a small but pre-trained model based on T5, and the second method is based on zero-shot query generation using GPT-4. For query generation using T5, we applied the BeIR query generation\footnote{\url{https://huggingface.co/BeIR/query-gen-msmarco-t5-large-v1}} model that uses a T5-based model pre-trained on the MS MARCO Passage dataset, and generated $100$ queries per passage. We also generated one query per passage using the GPT-4 based approach. 

The 2023 query sample for participants comprised 200 human queries, 250 T5 queries and 250 GPT-4 queries. The 200 queries were selected from a set of held-out MS MARCO queries that were not used in corpus construction, as done for the first time in the 2022 track. The synthetic queries were determined by sampling 250 T5 queries from different seed passages, and including the set of GPT-4 queries from the same 250 seed passages. The T5 queries were sampled to match a target sample of 250 NIST 2022 qrels. The NIST qrels were passage labels 3 and 4, so considered relevant by human relevance assessors, and sampled at random. We used two statistics for stratification, the query length and number of query words that lexically match a passage word (with no stopword removal or stemming). So, for example, if our target human qrel had a 7 word query with 3 of those occurring in the passage, we would pick a T5 query-passage pair with a 7-word query with 3 of those occurring in the passage. This gave us 250 T5 query-passage pairs with exactly the same query length and lexical overlap statistics as the 2022 NIST qrel target sample.


For the queries that were provided to the NIST assessors, the assessors removed the queries that do not look reasonable, that contain too few or too many relevant documents as these queries tend to be noisy or not very informative for evaluation purposes. Amongst the 48 T5-generated queries, 13 ($27.1\%$) of them were selected and amongst the 49 queries generated using GPT-4, 18 ($36.7\%$) of them were selected to be included in the test collection. For real queries, out of the 147 queries provided to the assessors, 51 ($34.7\%$) of them were selected. 

Table~\ref{tbl:queries_statistics} shows the total number of queries included in the test collection for each query type, together with average, min and max query length for each category. It can be seen that queries generated using GPT-4 tend to be much longer than the queries generated using T5 and the real queries and in general queries generated using T5 tend to be shorter than the other query types. 

For all query types, depth-10 pooling was used to select the documents to be judged by the NIST assessors. For each query type, Table~\ref{tbl:numrels_perquery} shows the average number of documents per query for each relevance grade. It can be seen that synthetic queries contain much fewer relevant documents (documents with relevance grade greater than zero) compared to the real queries (120.1 documents of relevance grade greater than zero for real queries vs. 71.76 documents for queries generated by T5 and 77.87 documents for queries generated by GPT-4.) Also, pools constructed using queries generated by T5 tend to contain significantly more non-relevant documents compared to the rest of the query types, suggesting that these queries may be more difficult than the other queries. 



\begin{table}
    \centering
    \caption{Statistics of queries per query type}
    \label{tbl:queries_statistics}
        \begin{tabular}{lccccc}
            \toprule
            \textbf{} & \textbf{All} & \textbf{Real} & \textbf{T5 Generated} & \textbf{GPT-4 Generated} \\
            \midrule
             No.~of Queries & 82 & 51 & 13 & 18 \\
             Avg.~Query Length & 6.84 & 5.76 & 5.69 & 10.72 \\
             Min Query Length & 2 & 2 & 4 & 6 \\
             Max Query Length & 15 & 14 & 8 & 15 \\
            \bottomrule
        \end{tabular}
\end{table}

\begin{table}
    \centering
    \caption{Average number of documents for each relevance grade for different query types.}
    \label{tbl:numrels_perquery}
        \begin{tabular}{cccccc}
            \toprule
            \textbf{Relevance Grade} & \textbf{All} & \textbf{Real} & \textbf{T5 Generated} & \textbf{GPT-4 Generated} \\
            \midrule
             0 & 169.09 & 159.31 & 213.30 & 164.88 \\
             1 & 53.31 & 64.15 & 31.23 & 38.55 \\
             2 & 27.54 & 31.60 & 19.46 & 21.88 \\
             3 & 22.31 & 24.35 & 21.07 & 17.44 \\
            \bottomrule
        \end{tabular}
\end{table}

\begin{figure}
  \center
  \begin{subfigure}{.49\textwidth}
    \includegraphics[width=\textwidth]{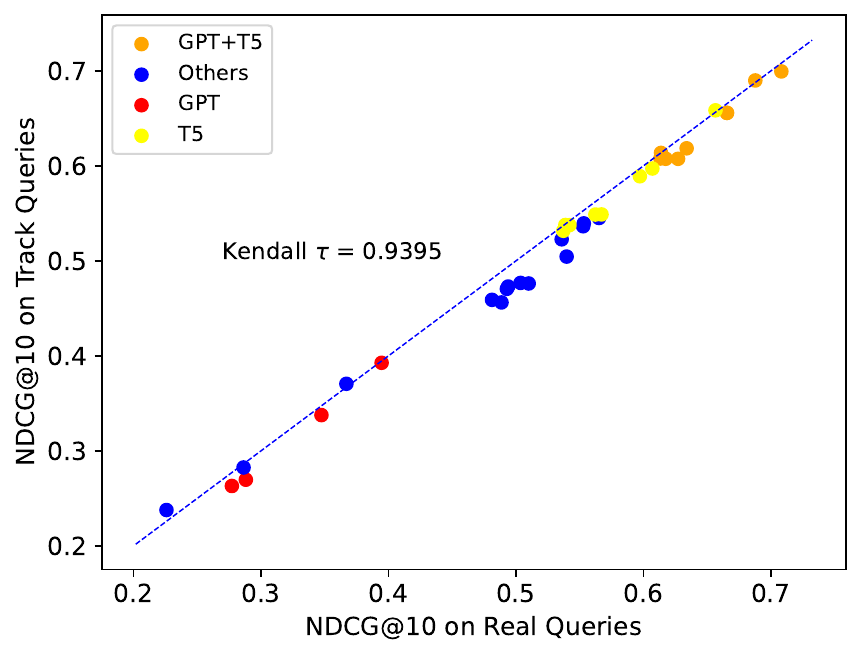}
    \label{fig:run_rank_real_Track}
  \end{subfigure}
  \hfill
  \begin{subfigure}{.49\textwidth}
    \includegraphics[width=\textwidth]{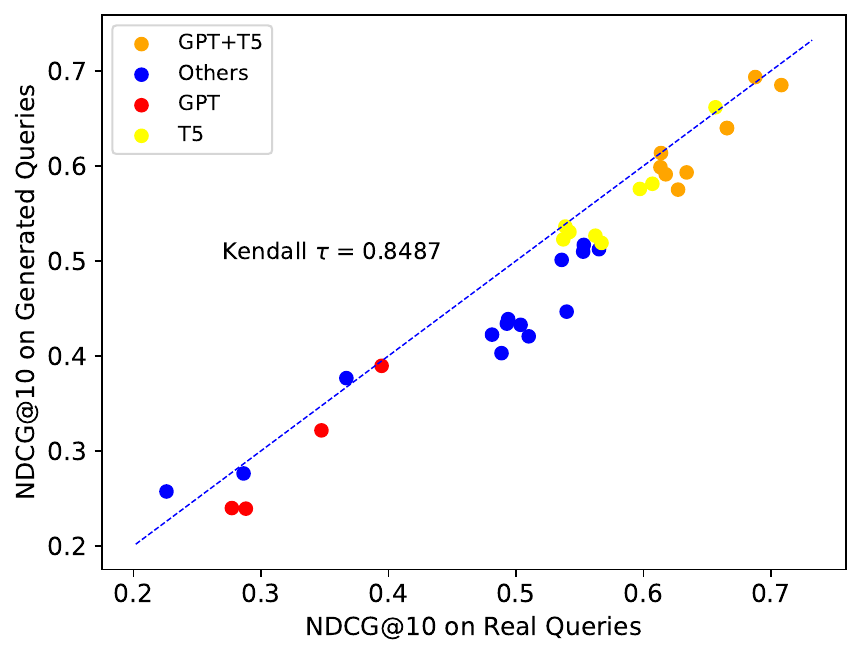}
    \label{fig:run_rank_real_Generated}
  \end{subfigure}
  \caption{
  Comparison of system performance on real queries versus (left) on all (real+synthetic) track queries, and (right) on synthetic queries.}
  \label{fig:synthetic-analysis-ndcg10}
\end{figure}

To compute the reliability of the final test collection which includes both real and synthetic queries, we compared the performance of systems using all track queries with performance of systems  solely using real queries. The left plot in Figure~\ref{fig:synthetic-analysis-ndcg10} shows that evaluation results obtained using our final test collection is very close to the evaluation results on real queries. The right plot in the figure shows how the performance of systems using synthetically generated queries (both T5 and GPT-4 based) compare with system performance on real queries. Both plots include line $y=x$ and report the Kendall's $\tau$ values. It can be seen that synthetic queries and real queries show similar patterns in terms of evaluation results and system ranking.

One potential issue with using synthetic queries in test collection construction is the possible bias the synthetic queries may exhibit towards systems that are based on a similar approach (similar language model) to the one that was used in query generation (e.g., queries generated using T5 might favour systems that are based on T5). In order to analyse the possible bias, we categorised the runs submitted to the track based on the approach they use, resulting in four different system categories: systems based on GPT, T5, GPT + T5 (i.e., a combination of GPT and T5), and others (traditional methods such as BM25, or any model that does not use either GPT or T5). The right plot in Figure~\ref{fig:synthetic-analysis-ndcg10} shows that while synthetic queries slightly underestimate the performance of models that do not use GPT or T5, this does not seem to have much effect in system ranking. 

\begin{figure}
  \center
  \begin{subfigure}{.49\textwidth}
    \includegraphics[width=\textwidth]{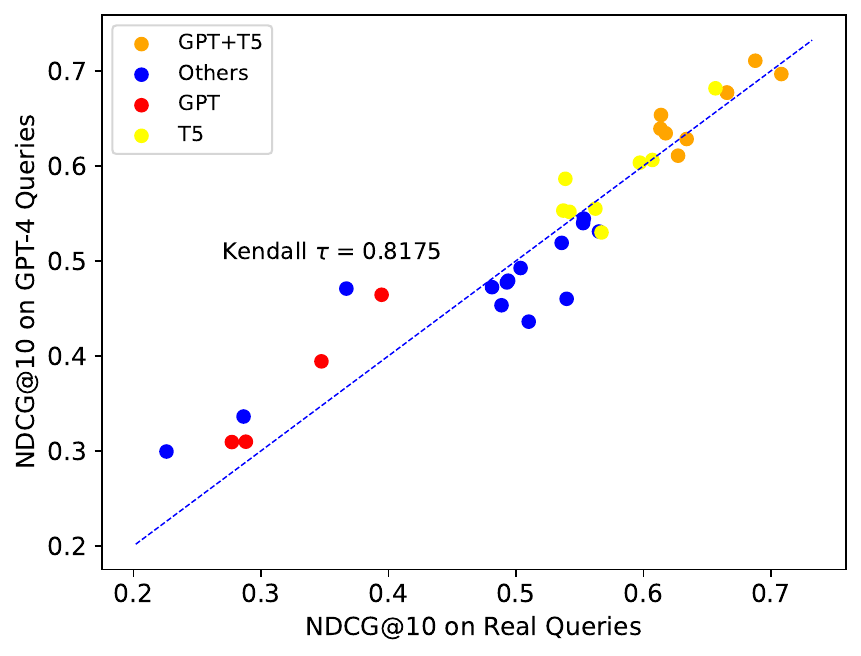}
    \label{fig:run_rank_real_GPT-4}
  \end{subfigure}
  \hfill
  \begin{subfigure}{.49\textwidth}
    \includegraphics[width=\textwidth]{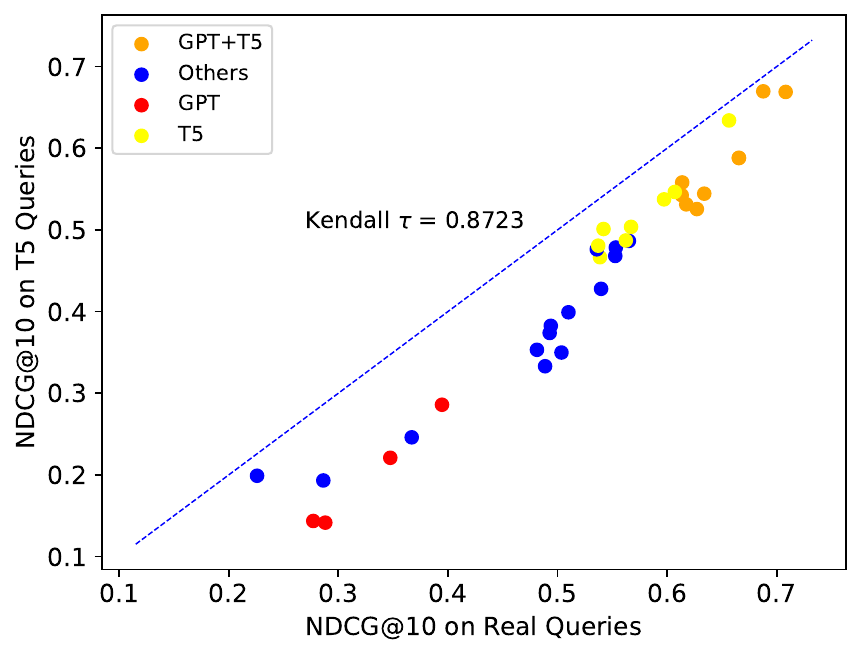}
    \label{fig:run_rank_real_T5}
  \end{subfigure}
  \caption{Comparison of system performance on real queries versus (left) on queries generated using GPT-4, and (right) on queries generated using T5.}
  \label{fig:synthetic-analysis-gptvst5}
\end{figure}

To further analyse possible bias that might arise from a system using a similar approach as the one used in query generation, the left plot in Figure~\ref{fig:synthetic-analysis-gptvst5} shows how system performance computed on GPT-4 queries compare with system performance on real queries. It can be seen that queries based on GPT-4 slightly over-estimate the performance of systems based on GPT. As it can be seen in the right plot of the figure, queries generated using T5 exhibit almost no bias towards systems based on T5. The plot also shows that for all system types, T5 based queries tend to be more difficult than real queries.

Overall, our initial results suggest that test collections consisting of synthetically generated queries could be reliably used to evaluate system performance. However, more analysis is needed to validate this, which is left as future work.

\section{Conclusion}
\label{sec:conclusion}
This is the fifth and final year of the TREC Deep Learning track.
Our main goal was to generate another reusable dataset, using the new methods introduced last year, giving us a second round of data.
This means we repeated the updats that were first introduced last year. A set of harder queries (not used in corpus construction), focused judgement on the passage task, and passage deduplication.
This year we also introduced some additional synthetic queries. 
Our initial analysis shows that the synthetic queries based on T5 and GPT-4 can lead to similar evaluation outcomes in comparison to evaluating with our sample of queries from search engine users.
Deep learning models with large scale pretraining continued to outperform traditional retrieval methods, and single stage retrieval with deep models seems to gain some more ground this year.
This report summarizes our analysis of submitted runs and the observed (mostly positive) impact of the changes in the track this year on building a more complete and consequently more reusable test collections.

\bibliographystyle{plainnat}
\bibliography{bibtex}

\end{document}